\documentstyle[preprint,prb,aps]{revtex} 
\begin{document}
\draft

\title{First-Principles Investigation of 180$^\circ$
       Domain Walls in BaTiO$_3$}

\author{J.~Padilla, W.~Zhong, and David Vanderbilt}

\address{Department of Physics and Astronomy,
Rutgers University, Piscataway, NJ 08855-0849}

\date{September 21, 1995}
\maketitle

\begin{abstract}
We present a first-principles study of 180$^{\circ}$ ferroelectric
domain walls in tetragonal barium titanate.  The theory is based on
an effective Hamiltonian that has previously been determined from
first-principles ultrasoft-pseudopotential calculations.
Statistical properties are investigated using Monte Carlo
simulations.  We compute the domain-wall energy, free energy, and
thickness, analyze the behavior of the ferroelectric order
parameter in the interior of the domain wall, and study
its spatial fluctuations.  An abrupt reversal of the polarization
is found, unlike the gradual rotation typical of the ferromagnetic
case.
\end{abstract}

\pacs{77.80.-e, 77.80.Dj, 71.10.+x, 82.65.Dp}

\narrowtext

The cubic perovskites are among the most important examples of
ferroelectric materials.\cite{line}  Many undergo not just one, but
a series, of structural phase transitions as the temperature is
reduced.  These transitions occur as a result of a delicate balance
between long-range dipole-dipole interactions that favor the
ferroelectric state, and short-range forces that favor the
high-symmetry cubic perovskite phase.  Because of the anomalously
large Born effective charge of the atoms, the ferroelectric
transitions in the perovskites are very sensitive to electrostatic
boundary conditions.\cite{resta,zhong1}  As a consequence, domain
structure plays an important role in the ferroelectric transitions,
and a theoretical understanding of the domain walls is of great
interest.

Theoretical investigation of ferroelectric domain walls has been
much less extensive than for their ferromagnetic counterparts. The
strong coupling of ferroelectricity to structural and elastic
properties is problematic.  Previous theoretical investigations
have concentrated on a phenomenological level of description, using
Landau theory to study domain-wall thickness and energy.\cite{zhv,bula}
Simple microscopic models such as local-field theory have also been
used to identify the domain-wall structure and character.\cite{law}
Due to the limited experimental data available, this empirical work
has tended to be qualitative and oversimplified, and has thus not
been able to offer the accuracy needed for a deeper theoretical
understanding.

In this paper, we undertake a first-principles study of domain
walls in BaTiO$_3$.  To our knowledge, this is the first such study
in any ferroelectric material.  Using an {\it ab initio} effective
Hamiltonian developed previously to study the phase transitions of
BaTiO$_3$,\cite{wz2} we set up Monte Carlo (MC) simulations
to investigate the structure and energetics of 180$^\circ$ domain
walls of (100) orientation. In particular, the energy, free energy,
and thickness
of the wall are calculated. We also analyze the behavior of the
ferroelectric order parameter in the interior of the domain wall,
and study the fluctuations in the domain-wall shape.  Where
we can compare with previous work, we find our results in general
agreement with experimental\cite{th,merz,fousek} and
theoretical\cite{zhv,bula,law} reports.

Because only low energy distortions are important to the structural
properties, we work with an effective Hamiltonian written in terms
of a reduced number of degrees of freedom.\cite{wz2} The most
important degrees of freedom included are the $3N$ ``local-mode
amplitudes'' $u_{i\alpha}$ for site $i$ and Cartesian direction
$\alpha$.  A ``site'' is a primitive unit cell centered on a Ti
atom, and the ``local mode'' on this site consists of displacements
of the given Ti atom, its 6 nearest oxygen neighbors, and its 8
nearest Ba neighbors, in such a way that a superposition of a
uniform set of local-mode vectors ${\bf u}_i={\bf e}$ (independent
of $i$) generates the soft zone-center ferroelectric mode polarized
along $\hat e$.  We also include six degrees of freedom to
represent homogeneous strain of the entire system, and $3N$
displacement local-mode amplitudes $v_{i\alpha}$ that serve to
introduce inhomogeneous strains.  We thus reduce the number of
degrees of freedom per unit cell from fifteen to six, simplifying
the expansion considerably.

Since the ferroelectric transition involves only small structural
distortions, we represent the energy surface by a Taylor expansion
around the high-symmetry cubic perovskite structure, including up
to fourth-order anharmonic terms where appropriate.  The energy
consists of five parts: an on-site local-mode self-energy, a
dipole-dipole interaction, a short-range interaction between local
modes, an elastic energy, and a coupling between the
elastic deformations and the local modes.  The Hamiltonian is then
specified by a set of expansion parameters, which are
determined using highly accurate LDA calculations with Vanderbilt
ultrasoft pseudopotentials.\cite{vand1} The details of the
Hamiltonian, the first-principles calculations, and the values of
the expansion parameters have been reported elsewhere.\cite{wz2}
This scheme has been successfully applied to single-domain
BaTiO$_3$ to predict the phase transition sequence, transition
temperatures, and other thermodynamic properties with good accuracy.

The phase transition sequence for BaTiO$_3$ is cubic to tetragonal
to orthorhombic to rhombohedral as temperature is reduced.  We
focus on the tetragonal phase, since it is the room-temperature
phase, and the best studied experimentally.  We adopt the
convention that the polarization, and thus the tetragonal $c$-axis,
are along $\hat z$.  In this phase, two kinds of domain walls,
90$^{\circ}$ and 180$^{\circ}$ walls, are possible.\cite{jano}
The notation refers to the angle between polarization vectors in
adjacent domains.  We choose the 180$^{\circ}$ domain wall for
this study because of preliminary indications of a simple structure
and narrow width.\cite{zhv}
Because it is energetically unfavorable to form domain walls carrying
net bound charge, 180$^{\circ}$ domain walls are restricted to
lie parallel to the polarization.  Earlier work has indicated that the
180$^{\circ}$ domain wall of $(100)$ orientation has much lower
energy than for other, e.g. $(110)$, orientations.\cite{law}
Thus, we focus on the $(100)$ domain wall, which we take to lie in
the $y$-$z$ plane.

Previous work has indicated that the width of the 180$^\circ$ domain
wall is very narrow, of the same order of magnitude as the lattice
constant.\cite{zhv,law,th}
For a very narrow domain wall, our choice of local mode
(Ti-centered as opposed to Ba-centered) may introduce some bias.
The point is that the sharpest domain wall that can be constructed
is one for which the local-mode vectors ${\bf u}_i$ are constant
except for a sudden sign reversal from one plane of sites to the
next.  For the Ti-centered choice of local modes, this represents a
Ba-centered domain wall, for which the atomic displacements have
odd symmetry across (and vanish on) the central Ba plane.
Conversely, for a Ba-centered choice of local mode, the sharpest
domain wall is Ti-centered, vanishing on a central plane of Ti
atoms.  In order to determine which of these scenarios is the more
realistic, we constructed $4\times 1 \times 1$ supercells
(containing 20 atoms and two domain walls) corresponding to each of
the above scenarios, using a mode amplitude taken from the average
equilibrium structure of the MC simulations (very close to the
experimental structure).  We then performed LDA calculations to
compare the energies of the two structures.  We find the
Ba-centered and Ti-centered walls constructed in this way have
energies of 6.2 erg/cm$^2$ and 62.0 erg/cm$^2$, respectively.
Thus, a sharp Ti-centered domain wall appears very unfavorable, and
it is clearly best to use a Ti-centered local mode as we have
done.  We note that the effective Hamiltonian reproduces the energy
of this sharpest Ti-centered wall to within 1\% of the LDA result
(not surprisingly, since configurations of a similar kind were
included in the fitting\cite{wz2}).

We study the structure and energetics of the domain walls using
Metropolis MC simulations.\cite{binder}  The degrees of
freedom are the vectors ${\bf u}_i$ and ${\bf v}_i$ for each
site $i$ of a $4L\times L\times L$ supercell, and the six
homogeneous strain components.  As discussed below, the supercell
is arranged to contain two domains, each roughly of size
$2L\times L\times L$, with domain walls normal to $\hat x$.
Periodic boundary conditions on the
supercell are assumed. Since all energy contributions (except for
the dipole-dipole coupling) are local, we choose the single-flip
MC algorithm.  We make a trial move of variables at one
site, check acceptance, make the change if accepted, and go on to
the next site.  One Monte Carlo sweep (MCS) constitutes one entire
pass through the system in this manner.

To generate a reasonable starting configuration, we equilibrate an
$L\times L \times L$ supercell at a high temperature ($T>$400 K)
in the cubic phase, and then cool it down slowly, allowing it to
relax for 20,000 MCS's at each temperature step.  We stop the
cooling when the tetragonal phase is reached, in which the polarization
vector averaged over the simulation cell points along one Cartesian
axis.  (As reported in Ref. \onlinecite{wz2}, this phase corresponds
to the temperature range from 230-290 K in our calculation, while
the actual experimental range is 278-403 K.\cite{mitsui})  If the
polarization is not along $+\hat z$, we rotate the structure to
make it so.  We then copy the structure four times along the
$x$-axis, with the polarization reversed to $-\hat z$ for two of
them, as shown in Fig.~\ref{f_cell}.

The supercell is thus constructed to contain two periodic 180$^{\circ}$
domain walls perpendicular to the $(100)$ direction.  This structure is
initially equilibrated for 2,000 MCS's, and then thermodynamic
averages are constructed from runs of 40,000 MCS's.
Since a supercell with domain walls has a higher energy than one
with just a single domain, we have found that the domain walls
sometimes spontaneously disappear during long simulations.  To
prevent this, we fix the $z$-components of the $\bf u$
vectors in the central two layers in each domain during the
simulations.  Since this structure should be very close to equilibrium
and the constrained components are far from the domain wall, we think
the effect on our results is negligible.

Fig.~\ref{f_pol} shows a snapshot of the polarization vector components
averaged over $y$-$z$ layers, $\overline u_x$, $\overline u_y$,
and $\overline u_z$, as a function of $x$, for $L=10$.  Several
qualitative features are immediately apparent.  First, the sharp
reversal of $u_z$ indicates that the domain boundary is indeed very
sharp, its width being on the order of a lattice constant.
Second, the other components $u_x$ and $u_y$ remain small throughout
the whole supercell, and their random fluctuations do not appear
to be correlated with the domain-wall position.  (The qualitative
difference between the fluctuations of $\overline u_x$ and
$\overline u_y$ with $x$ is an artifact of the averaging and of
the presence of strong longitudinal correlations.\cite{explan})
Thus, we find that the domain boundary entails a simple {\it reversal},
rather than a {\it rotation}, of the ferroelectric order parameter.

These behaviors are to be contrasted with the case of ferromagnetic
domain walls, where the magnetization vector typically rotates
gradually (on the atomic scale), keeping a roughly constant
magnitude.  This difference in behavior can probably be attributed
largely to the much stronger strain coupling in the ferroelectric
case.  For our BaTiO$_3$ geometry, for example, the entire
sample, including the interface, develops a tetragonal strain
along $\hat z$, imposed by the presence of domains polarized along
$\pm \hat z$.  This gives rise to a strong anisotropy
which will tend to keep the ferroelectric order parameter
from developing components along $x$ or $y$ in the interface region.
Thus, instead of rotating, the polarization simply decreases in
magnitude and reverses as we pass through the domain wall.
This absence of rotation of the polarization has been experimentally
verified for the case of the 90$^{\circ}$ domain wall
in BaTiO$_3$.\cite{yakunin}

We now turn to a quantitative analysis of our simulation results,
focusing on the domain-wall width, smoothness, and
energy.  We first estimate the domain-wall thickness $t$ as
follows.
For a string of sites along $x$ at a given value of
$(y,z)$ and on a given MCS, we identify the pair of sites
between which $u_z$ changes sign.  We then define
$t$ via the linear extrapolation $t/a=2u^{\rm spont}/\Delta u_z$,
where $a$ is the lattice constant, $u^{\rm spont}$ is the
spontaneous polarization deep in a domain, and $\Delta u_z$ is
the change of $u_z$ between the two interface sites.
Finally, we average over $(y,z)$ points and over MCS's
to get an average value of $t$.
The value of $t$ estimated in this way is 1.4 unit cells,
or 5.6 \AA. This is in reasonable agreement with empirical
theoretical estimates of 6.7 \AA\ (Ref. \onlinecite{bula}), and
experiments which place an estimated upper bound of 50 \AA\
(Ref. \onlinecite{th}).

To analyze the smoothness of the domain wall, we Fourier transform
the polarization $u_z$ as a function of the $x$ coordinate for each
$(y,z)$ point, and retain only the first three terms in the expansion.
This is an effective way to smooth the data while keeping the most
useful information.  The positions of the two domain walls in the
supercell, denoted by $X_1$ and $X_2$, are identified with the
values of $x$ at which the Fourier-smoothed $u_z$ changes sign.
In this way we obtain $X_1(y,z,\tau)$ and $X_2(y,z,\tau)$, where
$\tau$ labels the MCS.

In Fig. \ref{f_x1}, we show the probability distribution of $X_1$
for $L=10$ and for a run of 40,000 MCS's at 260 K.
The solid line is a histogram of the values of
$X_1(y,z,\tau)$, while the dashed line is a
histogram of $y$-$z$ planar averages $\overline X_1(\tau)$.
A comparison of the two curves shows that the spatial fluctuations
of the domain-wall position are much smaller than its ensemble
fluctuations.  From the solid line,
we see that the $X_1$ values have a typical standard deviation
of between 1 and 2 lattice constants.  Other runs indicate
that this result is not very sensitive to system size.
So, we can conclude that the domain
walls are relatively smooth.  We can further separate the
contributions to these fluctuations coming from the $y$- and
$z$-directions.  It is found that the fluctuations along the
$z$-direction (i.e., along the polar direction) are about 40\% smaller
than along the $y$-direction.  The sign of this
result was to be expected, since
the shape of the domain wall should be such as to minimize the surface
charge $\Delta\bf P \cdot \bf n$ that develops on it. Here, $\Delta\bf
P$ is the change of the polarization vector across the domain wall, and
$\bf n$ is the unit vector normal to the wall.

Finally, we turn to an estimate of the domain-wall formation energy.
Because of the periodic boundary conditions imposed on our system,
there are no surfaces to give rise to a depolarization energy.  Thus,
the domain-wall energy $E_w$ can be calculated from the difference
between the energy of the $4L\times L \times L$ supercell with and
without domain walls. This difference is small, but because the
correlation time of the system (far from the transition) is quite short
(20 MCS's), a sufficiently long simulation is capable of reducing
the statistical errors in $E_w$ to an acceptable level.  The
calculated domain-wall energies are shown in Table \ref{table1}.
The reported values have a statistical uncertainty of about 4\%.
Simulations for two lattice sizes and temperatures are reported. We
can see that our results are well converged with respect to system
size. Because of the large increase of the correlation time near
the transitions, it has proven difficult to give accurate values for
$E_w$ at other temperatures.

Our calculated value of $E_w$=16 erg/cm$^2$ for the domain-wall {\it
energy} is, however, probably not the proper quantity to compare with
experimentally derived values.  Instead, we should compute a {\it
free energy}, $F_w$, which includes entropic contributions from
fluctuations of the ferroelectric order parameter in the vicinity of
the domain wall.  A glance at Fig.\ \ref{f_pol}, which shows
considerable fluctuations, suggests that such contributions are likely
to be important.

We have estimated the domain-wall free energies $F_w$ using an
adiabatic switching technique, as follows.  First, we start with
an equilibrated $4L\times L\times L$ supercell containing two
domain walls, and for which the $z$-components of the $\bf u$
vectors in the central two layers in each domain are constrained
to preset values, as before.  We slowly reverse the values
of the constraint variables in the center of one of the domains
over the course of a 20,000-MCS simulation, making a small change
in the constraint variables every 10 MCS's, and compute the total
work done on the constraint variables.  If the simulation succeeds
in removing the two domain walls adiabatically, we can equate the
work done to twice the domain-wall free energy $F_w$.  By
comparing runs of from 20,000 to 30,000 MCS's, we find differences
in computed $F_w$ values of only about 10\%, which suggests that
the switching is indeed adiabatic.  The resulting computed values
of $F_w$ are about 4-5 erg/cm$^2$, or about 3-4 times smaller
than the $E_w$ values (and slightly smaller than the 6.2 erg/cm$^2$
reported above for the energy of the ideal Ba-centered wall).

Our result is consistent with previously published estimates of
the ``energy'' of the $(100)$ 180$^{\circ}$ domain wall, although
such previous values are rather scattered and inconclusive.
Previous experimental results of 10 erg/cm$^2$ and 3 erg/cm$^2$
(close to the phase transition) were given by Merz\cite{merz} and
Fousek and Safrankova,\cite{fousek} respectively.  On the
theoretical side, Bulaevskii\cite{bula} reported a value of 10.5
erg/cm$^2$ using a continuum Landau $p^6$ model, while
Lawless\cite{law} calculated an energy of 1.52 erg/cm$^2$ based on
a microscopic phenomenological model.
(Since the above estimates involve use of empirical models fit
to finite-temperature data, the ``energy'' values are probably
best interpreted as free energies.)

This investigation has opened several avenues for further studies.
Using the same method described here, it should be possible to
carry out a similar analysis for other types of domain walls,
e.g., $(110)$ 180$^{\circ}$ or 90$^{\circ}$ domain walls.
Bigger MC simulations on larger systems would allow the computation of
more precise values for the energy and thickness of the domain wall.

In summary, we have studied the properties of 180$^{\circ}$ domain
walls in BaTiO$_3$ using a first-principles based approach, by
applying Monte Carlo simulations to a microscopic effective
Hamiltonian that was fitted to {\it ab-initio} total-energy
calculations.  The simulations were carried out in the middle of
the temperature region of the tetragonal phase, relatively far from
the C--T and T--O transitions.  We confirm that the domain walls
are atomically thin, and that the order parameter does not rotate
within the wall.  We quantify the width, smoothness, and energetics
of these domain walls.  Our theoretical values of the wall width
and free energy are in reasonable agreement with previously
reported values, where available.

This work was supported by the Office of Naval Research under
contract number N00014-91-J-1184.


\begin{table}
\caption{Calculated domain-wall energies $E_w$ and free energies
$F_w$ as a function of simulation cell size $L$ and temperature
$T$. Statistical uncertainties are about 4\% for $E_w$ and 10\% for
$F_w$.
\label{table1}}
\begin{tabular}{cccc}
$T$ (K) &   $L$  & $E_w$ (erg/cm$^2$) & $F_w$ (erg/cm$^2$)   \\
\hline
250     &    8   &   15.8             &    4.6     \\
260     &    8   &   17.1             &    4.0     \\
250     &   10   &   15.6             &    5.0     \\
260     &   10   &   17.0             &    4.4     \\
\end{tabular}
\end{table}

\begin{figure}
\caption{
Schematic illustration of the simulation supercell arrangement.
\label{f_cell}}
\end{figure}

\begin{figure}
\caption{
Snapshot of the $y$-$z$ layer-averaged polarization-vector components
$\overline u_x$ (dotted line), $\overline u_y$ (dashed line),
and $\overline u_z$ (solid line), as a function of $x/a$ ($a$ is the
lattice constant), for the $40\times 10 \times 10$ lattice at 260 K.
\label{f_pol}}
\end{figure}

\begin{figure}
\caption{
Histograms of domain-wall positions for the $40\times 10 \times 10$
lattice at 260 K. Solid line, histogram of $X_1(y,z,\tau)/a$
values ($a$ is the lattice constant, $\tau$ labels a MCS);
dashed line, histogram of $y$-$z$ planar-average values
$\overline X_1(\tau)/a$.
\label{f_x1}}
\end{figure}

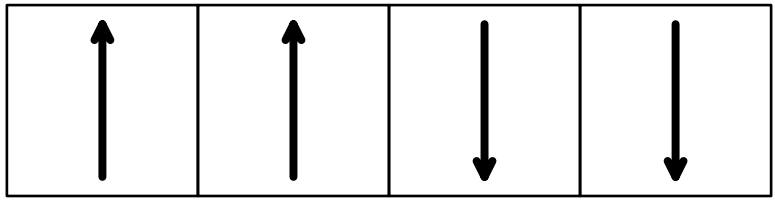
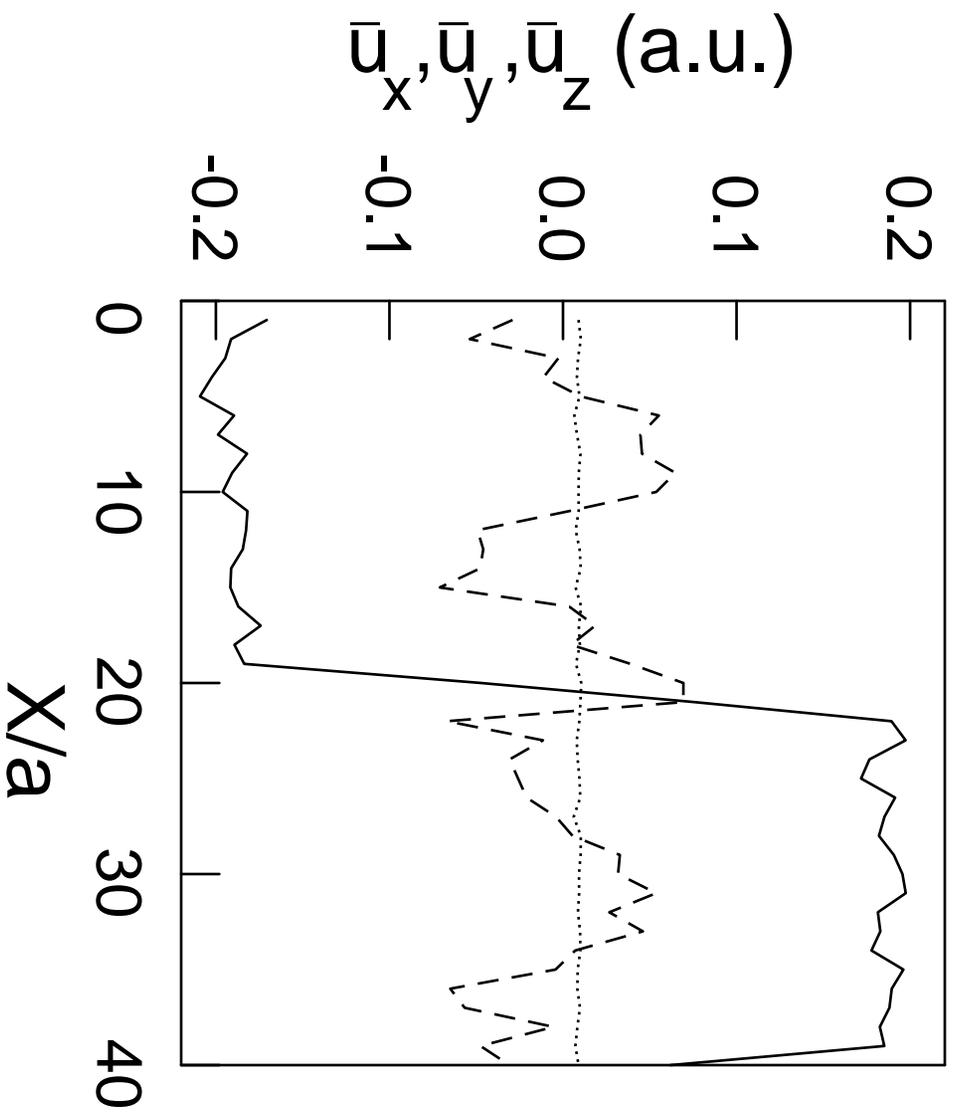
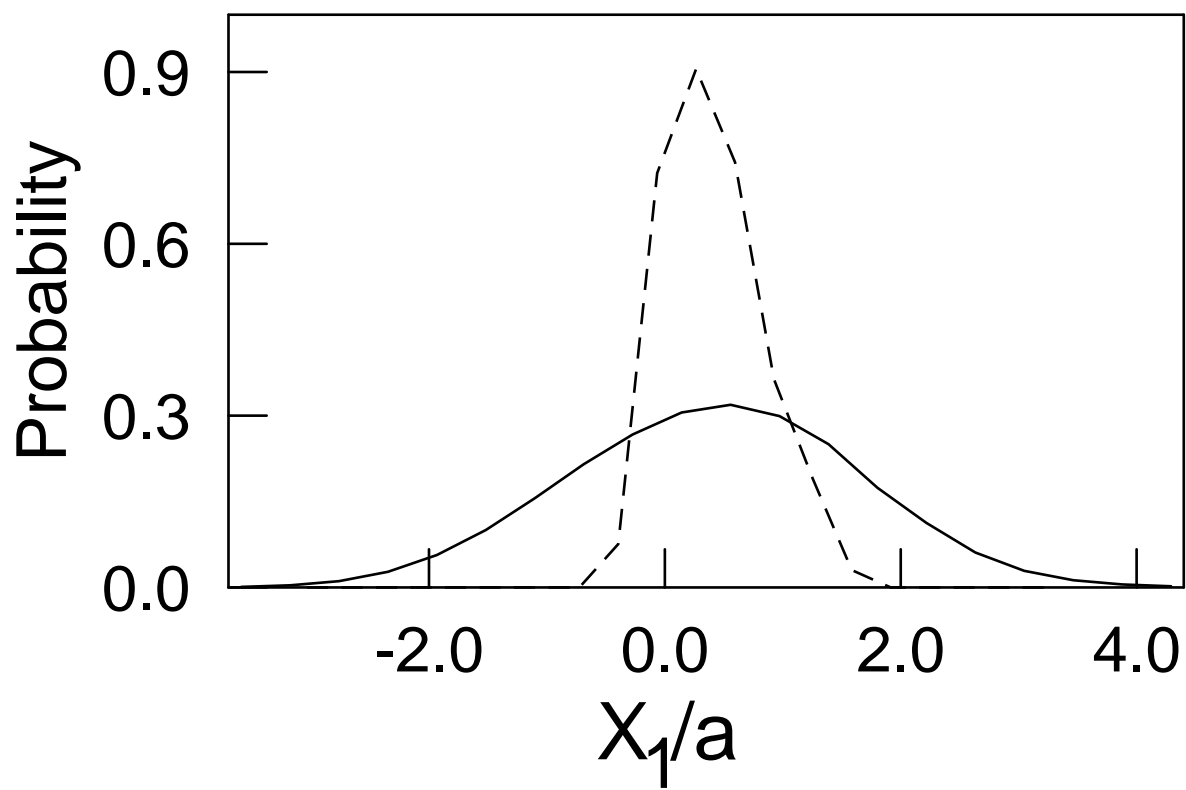
\end{document}